\documentclass[11pt]{article}
\usepackage{hyperref}
\usepackage{graphicx}
\pdfoutput=1

\begin{document}
\title{Turbulence-Flame Interaction}
\author{Brock Bobbitt$^a$ \and Bruno Savard$^b$ \and Guillaume Blanquart$^a$\\
\\\vspace{6pt} $^a$Department of Mechanical Engineering, \\ California Institute of Technology, Pasadena, C{A} 91125, USA\\
\\\vspace{6pt} $^b$Graduate Aerospace Laboratories, \\ California Institute of Technology, Pasadena, C{A} 91125, USA}
\maketitle

\begin{abstract}
A fluid dynamics video was created using data from a Direct Numerical Simulation (DNS) of a premixed {\it n}-heptane flame at high Karlovitz number. The magnitude of vorticity and progress variable(a monotonically increasing variable through the flame) illustrate the turbulence-flame interaction.
\end{abstract}
\section{Introduction}


The premixed turbulence-flame interaction, especially at the transition between the thin reaction zones and the broken reaction zones, contains many complex phenomenon and the coupling of fluid mechanics and chemistry. Direct Numerical Simulations (DNS) at high Karlovitz number(defined as the ratio of the chemical time scale to smallest turbulent time scale) and employing various fuels provide important information for the purposes of describing and modeling the behavior. The video displays data obtained from a DNS of a premixed {\it n}-alkane flame at a Karlovitz number corresponding to the transition between the thin reaction zone and broken reaction zone using a low-Mach number Navier-Stokes solver \cite{NGA}. 

\section{Video}
The video shows (1) the transition of turbulence through the flame front, and (2) the effects of turbulence on the flame front. The video contains images of variable light emission corresponding to the magnitude of the quantity displayed. The quantities displayed are the magnitude of (1) vorticity, (2) vorticity normalized by its time-averaged maximum at each cross-section (perpendicular to the mean flow), and (3) the progress variable. The scales of turbulence are captured illustrating how turbulence is transformed as it goes through the flame front. Similarly, the flame front is distorted by turbulence and is distinctly more diffuse than a laminar flame front.
\section{Conditions and parameters}
The fuel used is {\it n}-heptane ({\it n}-$C_7H_{16}$) with unburnt conditions of $p=1$ atm, $T_u=298$ K, and $\phi=0.9$. In the simulation, the chemistry is reduced to the use of a progress variable. 

The position of the flame is statistically steady, as the ends of the domain corresponding to an inflow of unburnt gases and an outflow of burnt gases. This allows for an indefinite simulation time, which gives the opportunity to obtain a well established and characterized flame. The parameters of the simulation are listed in Table~\ref{t:tab1}. Here, $\tau$ is the eddy turnover time and $t_{flow}$ is the flow time from the inflow to the outflow ($11L$). Since the ratio of $t_{\rm flow}$ to $\tau$ large, forcing of the turbulence is necessary. 
\begin{table*}[h]
\begin{center}
  \begin{tabular}{ p{6cm}  c}
  
    \hline\noalign{\smallskip}
    Height/width $L$ (m) & 2.33X10$^{-3}$ \qquad\\\noalign{\smallskip}
    Aspect ratio & 11 \qquad\\\noalign{\smallskip}
    Grid & 128 X 128 X 1408 \qquad\\\noalign{\smallskip}
    $l/l_F$ & 1.02 \qquad\\\noalign{\smallskip}
    $u^\prime/s_L$ & 21.4 \qquad\\\noalign{\smallskip}
    $Ka$ & 98 \qquad\\\noalign{\smallskip}
    $Re_t$ & 190 \qquad\\\noalign{\smallskip}
    $Re_\lambda$ & 53.4 \qquad\\\noalign{\smallskip}
    $\tau$ (s) & 1.07x10$^{-4}$ \qquad\\\noalign{\smallskip}
    $t_{\rm flow}/\tau$ & $\approx$100 \qquad\\\noalign{\smallskip}
    \hline
  \end{tabular}
  \end{center}
\caption{Simulation parameters}
\label{t:tab1}
\end{table*}

\end{document}